\documentstyle[preprint,aps,psfig]{revtex}
\tightenlines
\def\beq{\begin{equation}}
\def\eeq{\end{equation}}
\def\bea{\begin{eqnarray}}
\def\eea{\end{eqnarray}}
\def\nnu{\nonumber}

\def\al{\alpha}

\def\kap{\kappa}
\def\lam{\lambda}
\def\om{\omega}

\def\Dta{\Delta}
\def\Lam{\Lambda}

\def\Om{\Omega}

\def\ptl{\partial}

\def\hf{{1\over2}}

\def\lp{\left(}
\def\rp{\right)}

\def\ham{{\cal H}}
\def\ket#1{|#1\rangle}

\def\tran#1#2{\langle#1|#2\rangle}

\def\mel#1#2#3{\langle#1|#2|#3\rangle}

\def\bH{{\bf H}}
\def\bJ{{\bf J}}

\def\xhat{\bf{\hat x}}
\def\zhat{\bf{\hat z}}

\def\Mn12{Mn$_{12}$}
\def\Fe8{Fe$_8$}

\def\hsc{{\ham_{\rm sc}}}
\def\tbt{{3\over 2}}
\def\baJ{\bar J}

\def\D0{\Delta_0}

\def\e0k{E_0^{(k)}}

\begin{document}
\draft

\title{Oscillatory Tunnel Splittings in Spin Systems: A Discrete
Wentzel-Kramers-Brillouin Approach}

\author{Anupam Garg$^*$}
\address{Department of Physics and Astronomy, Northwestern University,
Evanston, Illinois 60208}

\date{\today}

\maketitle

\begin{abstract}
Certain spin Hamiltonians that give rise to tunnel splittings that are
viewed in terms of interfering instanton trajectories, are
restudied using a discrete WKB method, that is more elementary,
and also yields wavefunctions and preexponential factors for the
splittings. A novel turning point inside the classically
forbidden region is analysed, and a general formula is obtained
for the splittings. The result is appled to the \Fe8 system. 
A previous result for the oscillation of the ground state splitting
with external magnetic field is extended to higher levels.

\end{abstract}
\pacs{75.10Dg, 03.65.Sq, 36.90+f, 75.45.+j}

\widetext

The magnetic properties of the molecular cluster
[(tacn)$_6$Fe$_8$O$_2$(OH)$_{12}$]$^{8+}$ (or just \Fe8 for short) are
governed by a Hamiltonian\cite{alb}
\beq
\ham  = -k_2 J_z^2 + (k_1 - k_2) J_x^2 - g\mu_B \bJ\cdot\bH, \label{ham} 
\eeq
where $\bJ$ is a dimensionless spin operator, $\bH$ is an
externally applied magnetic field, $J=10$,
$k_1 \approx 0.33$ K, and $k_2 \approx 0.22$ K. The zero-field Hamiltonian
has biaxial symmetry with easy, medium, and hard axes along $z$, $y$, and
$x$ respectively \cite{fn1}. Very recently, Wolfgang and Sessoli\cite{ws}
have seen a new effect in this system, viz., an oscillation in the
Landau-Zener transition rate between Zeeman levels, as a function of
the applied field along $\xhat$.
These oscillations reflect an oscillation in the underlying tunneling
matrix element between the levels in question, and are in this author's
view, the only unambiguous evidence to date for quantum tunneling of a spin of
such a large size in a solid state system. This effect is not seen, e.g.,
in a closely related \Mn12 cluster.

Oscillations as a function of $H_x$ in the ground state tunnel splitting
$\Dta$ of the Hamiltonian (\ref{ham}) were in fact predicted
earlier\cite{agepl}, on the basis of an instanton calculation. For
$\bH \| \xhat$, there are two instantons, with a complex action differing
by a Berry phase that sweeps through odd multiples of $\pi$ as $H_x$ is
varied, leading to a complete quenching of tunneling. In this view
the effect arises from
destructive intereference between spin trajectories \cite{ldg}. A
different perspective was provided in Ref. \cite{agprb}, by noting
that $\ham$ is invariant under a $180^\circ$ rotation about $\xhat$
when $H_z=0$. The oscillation is due to a symmetry-allowed crossing of
levels with different parity under the rotation.

Although the oscillations with $H_x$ are easily seen by direct numerical
diagonalization of $\ham$ (see Fig.~1 of \cite{agepl},
or Fig.~4 of \cite{ws}), it is of interest to understand
these features analytically. Since the spin $J$ is large, it is
natural to use the semiclassical, or $J\to\infty$ approximation.
To some extent, the instanton method already does this.
In this paper, we will try and make further progress 
using a discrete WKB method \cite{sg,pab,vs,agjmp}, which has several
advantages.  First, it is very difficult to find the next-to-leading
terms in the $J\to\infty$ asymptotic expressions for various
physical quantities (such as $\Dta$)
using instantons. Second, Wernsdorfer and Sessoli also see an
oscillatory rate in the presence of a dc field along $\zhat$ that
is such as to align the ground level in one well with an excited level
in the other. The splitting is now never perfectly quenched as
the symmetry of $\ham$ is destroyed.  We do not (although others may)
know how to solve this problem with instantons. Third,
with an eye to the future, the method provides
wavefunctions in addition to energies, which may be used to calculate
matrix elements of various perturbations \cite{agprb,agsea}
to Eq.~(\ref{ham})
that are present in the actual physical system, and thus study their
influence. 

Specifically, we will derive a general result [see Eq.~(\ref{Spl2})]
for the tunnel splitting between degenerate pairs of levels in a symmetric
problem when oscillations are present. Our result is expressed in terms
of two action integrals. In the course of doing this, we will encounter
a novel feature that does not arise in previous discrete WKB studies,
namely, a turning point in the classically forbidden region!
We will apply our result to the hamiltonian (\ref{ham}) for $H_z=0$,
focussing in detail on quenching fields for ground and excited state
splittings. The results for the latter are new.  The study of
imperfect quenching of tunneling that occurs when $H_z \ne 0$, because
the potential is then asymmetric, is much more involved, and will be
published separately.

Let us first briefly review the discrete WKB formalism \cite{pab}.
The starting point is to write Schr\"odinger's equation
in the $J_z$ basis. Let $\ham\ket{\psi} = E \ket{\psi}$,
$J_z \ket m = m\ket m$, $\tran{m}{\psi} = C_m$,
$\mel{m}{\ham}{m} = w_m$, and $\mel{m}{\ham}{m'} = t_{m,m'}$ ($m \ne m'$).
Then we have
\beq
\sum_{n \ne m} t_{m,n} C_n + w_m C_m = E C_m. \label{Seq}
\eeq
We assume that the matrix $t_{m,n}$ is real and symmetric,
$t_{m,n} = t_{n,m}$.
In the present problem, we need matrix elements that are off-diagonal
by 1 ($t_{m,m\pm 1}$) {\it and} by 2 ($t_{m,m\pm 2}$). This makes
Eq.~(\ref{Seq}) a recursion relation involving five terms, as opposed
to three terms in previous work. The physical idea is to view
Eq.~(\ref{Seq}) as a tight-binding model for an electron hopping on a
one-dimensional lattice, and use the approximation of semiclassical electron
dynamics. This would be exact if the matrix elements of $\ham$ were
constant with $m$, and will be systematically justifiable if they
are slowly varying with $m$. Formally, the latter
means that we can find functions $w(m)$, $t_1(m)$, and $t_2(m)$
of a continuous variable $m$, such that on the discrete eigenset of $J_z$,
\bea
w(m) &=& w_m, \label{wcont} \\
t_{\al}(m) &=& (t_{m,m+\al} + t_{m,m-\al})/2, \quad \al=1,2,
   \label{tcont}
\eea
and further, that if $m/J$ is regarded as a quantity of order $J^0$, then
$\dot w(m)\equiv dw/dm = O(w(m)/J)$, with similar restrictions on
${\dot t}_1(m)$ and ${\dot t}_2(m)$. For Eq.~(\ref{ham}), these conditions
are met if $J \gg 1$.

If $w_m$, $t_{m,m\pm1}$, and $t_{m,m\pm 2}$ were constant, the eigenstates
of $\ham$ would be states with $C_m = e^{iqm}$, and
$E=w+2t_1\cos q + 2t_2 \cos(2q)$. Now we seek a solution in the
form $C_m = e^{i\Phi(m)}$ with $\Phi = \Phi_0 + \Phi_1 + \Phi_2 + \cdots$,
where $\Phi_n = O(J^{1-n})$, and $\dot\Phi_n = O(\Phi_n/J)$. Then, one can
show \cite{pab} that up to terms of order $J^0$ in $\Phi$, the solution is
given as linear combinations of the form
\beq
C_m \sim {1 \over \sqrt{v(m)}}\exp\lp i\int^m q(m') dm'\rp,
    \label{Cwkb}
\eeq
where $q(m)$ is a local wavevector that obeys the eikonal or
Hamilton-Jacobi equation,
\beq
E = w(m) + 2t_1(m) \cos q + 2t_2(m) \cos(2q)
     \equiv \hsc(q,m), \label{hjeq}
\eeq
and $v(m)$ is the associated semiclassical electron velocity,
which obeys the transport equation
\beq
v(m) = \ptl \hsc/\ptl q = -2\sin q(m)(t_1(m) + 4 t_2(m) \cos q(m)).
     \label{vm}
\eeq

To talk of tunneling, we must first understand the classically
allowed and forbidden regions in the $m$ space.
As a function of $q$ for fixed $m$, the semiclassical Hamiltonian
$\hsc(q,m)$ can be viewed as a band energy curve, and its minimum and
maximum values define local band-edge functions $U_\pm (m)$. The classically
accessible region for any energy $E$ is thus defined by
$U_-(m) \le E \le U_+(m)$. [The first consequence of having five terms in the
recursion relation shows up here. In the three term case, the band edges
always occur at $q=0$ or $\pi$. Now, they can occur at values other than
these if $|t_1(m)/4t_2(m)| < 1$. These functions are sketched in Fig.~1 for
Eq.~(\ref{ham}) with $H_z = 0$. The minimum, $U_-$, is attained at $q=0$ for
$|m| \ge m^*$, and at $q\ne 0$ for $|m| < m^*$. The curve $U_-(m)$ is
smooth at $m = \pm m^*$, and the formula for $m^*$ is unimportant.] Thus,
for the energies $E_a$ and $E_b$ drawn in Fig.~1, the central region
is classically forbidden and allowed, respectively. We will focus
on states of the first type in what follows.

The next step is to derive a generalization of Herring's formula for
the tunnel splitting $\Dta$ for a pair of levels whose mean energy is $E$.
Proceeding in exact analogy with Ref.~\cite{ll} (see also \cite{vs}(c,d)
or \cite{agjmp}), we consider a solution $C_m$ to Eq.~(\ref{Seq}) with
energy $E$, that is (a) localized in the left well of $U_-(m)$, and decays
away from that well everywhere
including the region near $m=0$, and (b) normalized
to unit total probability. The behavior of this
solution near the right well need not be specified or examined too
closely. Up to an irrelevant over all sign, we find
\beq
\Dta = \cases{
        2\left[ t_{01} C_0(C_1 - C_{-1}) + t_{02}C_0(C_2 - C_{-2})
                + t_{-1,1} (C_1^2 - C_{-1}^2) \right],
                                & \\integer $J$, \cr
        2\ t_{-\hf,\hf} \Bigl( C^2_{\hf} - C^2_{-\hf}\Bigr)
         + 4\ t_{-\tbt,\hf} \lp C_{\hf}C_{\tbt} - C_{-\hf}C_{-\tbt}\rp,
                                & half-integer $J$. \cr}
 \label{Hform}
\eeq

To apply Eq.~(\ref{Hform}), we must find $C_m$ in the central region.
In principle the procedure is straightforward, and follows
conventional WKB. We first find $C_m$ in the allowed region, near
$-m_0$, and then use connection formulas to extend it into the
forbidden region. For a three term recursion relation, this is
done in Ref.~\cite{agjmp}. In the present case, we encounter a
new difficulty. To see this, we consider points at which $v(m)$
vanishes. At all such points, which may be called turning points,
the solution (\ref{Cwkb}) diverges, indicating a breakdown of
the WKB approximation. Let us now consider a point strictly inside
the classically allowed region in the $E$-$m$ plane. At such a point
$q$ is not an extremum of $\hsc$ for fixed $m$, i.e., $v(m)\ne 0$.
It is a simple corollary that the points $E=U_{\pm}(m)$ {\it are} turning
points, corresponding to $q =0$, $\pi$, or $\cos^{-1}(-t_1/4t_2)$.
These turning points are of the same physical character as those
in conventional WKB, and the $q=0$ or $\pi$ ones are the only ones
that arise with a three-term recursion relation.

For our five-term recursion, however, $v(m)$ can also vanish
if $\cos q = -t_1/4t_2$, even though $E \ne U_{\pm}(m)$. To see
how this can happen, we solve Eq.~(\ref{hjeq}) to get
\beq
\cos q(m) = {-t_1(m) \pm [t^2_1(m) - 4t_2(m) f(m)]^{1/2}
              \over 4t_2(m)}, \label{cosq}
\eeq
where $f(m) = w(m) - 2t_2(m) -E$. Thus, such a turning point may
arise when the discrimiant of the quadratic equation for
$\cos q(m)$ vanishes. Since, by exclusion, such points
must necessarily lie in a classically forbidden region, where
$q(m)$ is not real, it follows that they can only
arise in problems where $|t_1(m)/4t_2(m)| > 1$ for some $m$. This fact
and Eq.~(\ref{cosq}) then imply that at such a point, $\cos q$ changes
from real to complex, i.e., $q$ changes from pure imaginary to
complex, and the wavefunction accordingly changes from an exponential
decay with one sign to a decay with an oscillating sign.

Since WKB breaks down at the forbidden region turning points, we
need connection formulas at these points just as for ordinary ones \cite{pab}.
We will publish the derivation of these formulas elsewhere, and here
we only give the result. Let the discriminant in
Eq.~(\ref{cosq}) vanish at $m=m_c$, and let $\cos q$ be real
for $m < m_c$, and complex for $m > m_c$. It is convenient to
define $q(m) = i\kap(m)$ with $\kap >0$ in the region $m < m_c$, and
to write $s(m) = -i v(m)$ everywhere. (This definition renders $s(m) > 0$
for $m > m_c$.) We consider the decaying WKB
solution in the region $m < m_c$:
\beq
C_m = {A \over 2 \sqrt{s(m)}}
      \exp \lp -\int\limits_{m_c}^m \kap(m') dm' \rp, \quad m <m_c,
  \label{Cleft}
\eeq
where $A$ is chosen to be real.  For $m > m_c$, 
we must consider linear combinations of the type (\ref{Cwkb}), with
two choices for $q(m)$ which we write as
\beq
q_{1,2}(m) = i\kap(m) \pm \chi(m). \label{kapchi}
\eeq
For the solution to continue decaying, we must still have $\kap>0$,
and we also choose $\chi >0$. Then, both $\kap(m)$ and $\chi(m)$ have
a kink at $m=m_c$.
We further define $s_{1,2}(m) = -iv(q_{1,2}(m))$ via Eq.~(\ref{vm}),
so that $s_2 = s^*_1$. The WKB solution which connects to (\ref{Cleft})
is then given by
\beq
C_m = {\rm Re} {A \over \sqrt{s_1(m)}}
      \exp \lp i\int\limits_{m_c}^m q_1(m') dm' \rp, \quad m > m_c.
   \label{Crite}
\eeq
Note that this is explicitly real, as the reality of Eq.~(\ref{Seq})
requires. Also, Eqs.~(\ref{Cleft}) and (\ref{Crite}) only hold for
$|m_c - m| \gg J^{1/3}$. The connection formula for the growing
solution is similar, but is not needed for our present purpose.

The result (\ref{Hform}) is exact, but does not reveal the physically
important barrier penetration factor. To remedy this, we substitute
Eq.~(\ref{Crite}) in Eq.~(\ref{Hform}).  
We consider a situation as in Fig.~1, with minima in
$U_-(m)$ at $\pm m_0$, and forbidden region turning points at
$\pm m_1$. The key, clearly, is to simplify Eq.~(\ref{Crite}) in the
region $ |m| < m_1$. To this end, we substitute Eq.~(\ref{kapchi}) for
$q$ in Eq.~(\ref{cosq}) and separate the real and imaginary parts. This
yields
\bea
\cosh\kap \cos\chi &=& -t_1/4t_2, \label{chcos} \\
\sinh\kap \sin\chi &=& (4t_2f - t_1^2)^{1/2}/4t_2. \label{shsin}
\eea
Using these results, it follows that
\beq
s_1 = 8t_2(m) \sinh\kap(m) \sin\chi(m) \sin q_1(m). \label{s1m}
\eeq
We now specialize to the case of integer $J$; the other is
similarly analysed, and yields the same result, Eq.~(\ref{Split})
below. For Eq.~(\ref{Hform}), we only need $C_m$ for $m= 0$,  $\pm 1$,
and $\pm 2$. The variations in $\kap$, $\chi$, $t_1$, $t_2$, and
$q_1$ between
these points may be ignored as they are of of order $J^{-2}$. Hence, to
sufficient accuracy one may write (for $|m| < 2$),
\beq
C_m = {\rm Re} A_2 { e^{i(\Om + m q_{10})}\over \sqrt{\sin q_{10}}},
   \label{Cmid}
\eeq
where $\Om = \int_{-m_1}^0 q_1(m')dm'$, and
$A_2 = (8t_{20} \sinh\kap_0\sin\chi_0)^{-1/2}A$. The suffix 0 denotes
quantities evaluated at $m=0$; thus $q_{10} = q_1(0)$, $\kap_0 = \kap(0)$,
etc. To the same accuracy as Eq.~(\ref{Cmid}) one may
write $t_{01} = t_1(0)$, and $t_{02} = t_{-1,1} = t_2(0)$ in
Eq.~(\ref{Hform}). If we use Eq.~(\ref{chcos}), and write every thing
in terms of $t_{20}$, $\kap_0$, and $\chi_0$, then a certain amount of
algebra leads to
\bea
\Dta &=& \hf A^2 (e^{2i\Om} + e^{-2i\Om^*}) \nnu \\
     &=& A^2 \exp \lp -\int\limits_{-m_1}^{m_1} \kap(m') dm' \rp
              \cos\lp \ \int\limits_{-m_1}^{m_1} \chi(m') dm' \rp.
     \label{Split}
\eea
The cosine factor clearly shows the possibility of oscillations.

The next step is to match the WKB wavefunction (\ref{Cleft}) in the
ordinary decaying region to the wavefunction in the allowed region.
It is plain that $A$ will contain an additional barrier penetration factor
$\exp(-\int_{m_1}^{m_t} \kap(m) dm)$, where $\pm m_t$ are the ordinary
turning points. Omitting the details of the  
calculation, which are very much like those in conventional WKB, we
find that for the $n$th pair of levels, provided $n \ll J$,
\beq
\Dta = {2 \om_0 \over \pi} g_n
      \exp \lp -\int\limits_{-m_t}^{m_t} \kap(m') dm' \rp
              \cos\lp \ \int\limits_{-m_1}^{m_1} \chi(m') dm' \rp,
     \label{Spl2}
\eeq
where $\om_0$ is the small oscillation frequency in the wells near
$\pm m_0$, and $g_n = \sqrt{2\pi} {\bar n}^{\bar n} e^{-\bar n}$ 
(with $\bar n = n + \hf$). It need hardly be said that $m_t$ and $m_1$
depend on the energy and hence on $n$.

Equation (\ref{Spl2}) is a general formula for the splitting in the
presence of interference effects. As opposed to an ``exponentially
accurate" calculation which gives an asymptotically correct result
for $\ln \Dta$ as $J \to\infty$, it is correct for $\Dta$ itself.

We now apply Eq.~(\ref{Spl2}) to the Hamiltonian Eq.~(\ref{ham}). The
problem is now merely one of quadrature, so we will focus only on the
cosine factor. (The full expression for the non-oscillatory
part of $\Dta$ including the exact prefactor is exceedingly lengthy and 
unilluminating. A partial result for the WKB exponent, or Gamow
factor may be found in Ref.~\cite{agprb2}.) In doing the quadratures,
the first step is to find $w(m)$, $t_{\al}(m)$, etc. Here, any 
function that reproduces the first two terms in a series in $1/J$
is adequate, since Eq.~(\ref{Cwkb}) represents only the two leading
terms in $\Phi(m)$. We define $\baJ = J + \hf$, $\mu = m/\baJ$,
$H_c = 2k_1 J/g\mu_B$, $h_x = J H_x/\baJ H_c$, $\lam = k_2/k_1$, and
measure all energies in units of $k_1\baJ^2$. Then,
$t_1 = -h_x(1-\mu^2)^{1/2}$, $4t_2 = (1-\lam)(1-\mu^2)$,
$2w = (1+\lam)(1-\mu^2)$, and $\om_0 = 2[\lam(1-h_x^2)]^{1/2}/\baJ$.
Further, let us denote the argument of the cosine in
Eq.~(\ref{Spl2}) by $\Lam$. The turning point $\mu_1(E)$ is given by
\beq
\mu_1^2(E) = 1 - {h_x^2 \over 1-\lam} - {E \over \lam}.
    \label{m1tp}
\eeq
Secondly, from Eq.~(\ref{shsin}), we see that
$\chi \sim (\mu_1(E) - \mu)^{1/2}$, so that to relative order
$1/J$, we may write
\beq
\Lam = 2\baJ \int\limits_0^{\mu_{10}}
       \lp \chi(\mu,E=0) +
             E{\ptl\chi \over \ptl E}\bigg|_{E=0} \rp d\mu,
     \label{Lam}
\eeq
where $\mu_{10} = \mu_1(0)$, and $E = (n+\hf)\om_0$ for the
$n$th pair of levels. At $E=0$, we have
$\cos^2\chi = [(1-\mu_{10}^2)/(1-\mu^2)]$,
$\ptl\chi/\ptl E = -\cot\chi/2(1-h_x^2 - \mu^2)$. Doing the
integrals, one obtains
\beq
\Lam = \pi J\lp 1 - {H_x \over \sqrt{1-\lam}H_c} \rp
              - n\pi.
   \label{Lam2}
\eeq
The result for $n=0$ is the same as in \cite{agepl}), while
for $n \ne 0$ it is new. To order $1/J$, the vanishing points
for higher pairs are the same as those for the lowest one.
It must be remembered, however, that since we demanded
$\chi >0$, one must have $\Lam > 0$. (Finding $\Lam < 0$
means that the oscillatory forbidden region has disappeared.)
Thus the highest-field level crossing is successively eliminated
as $n$ increases, and (including zero and negative values)
there are $2(J-n)$ fields in all where $\Dta$ vanishes.

\acknowledgments
This work is supported by the NSF via grant number DMR-9616749.

\begin{figure}
\caption{Band-edge functions $U_{\pm}(m)$. For energy $E_a$,
$m_t$ and $m_1$ are turning points, the latter in the classically
forbidden region.}
\end{figure}


\begin{references}
\bibitem{fn1} We omit fourth order terms that are necessary for a
quantitative description of the experiments, but do not alter the
essential concepts discussed here.
\end{references}
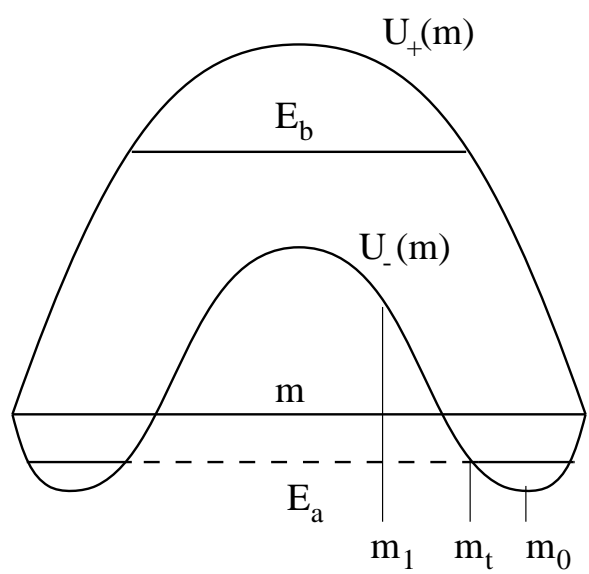
\end{document}